\documentclass[aps,prc,twocolumn,superscriptaddress]{revtex4-1}

\usepackage{graphicx}
\usepackage{multirow}
\usepackage{float}
\usepackage{dcolumn}
\usepackage{bm}
\usepackage{dcolumn}
\newcolumntype{d}[1]{D{.}{.}{#1}}
\begin{document}


\title{Half-Lives of Neutron-Rich $^{128-130}$Cd}


\author{R.~Dunlop}
\email[]{rdunlop@uoguelph.ca}
\affiliation{Department of Physics, University of Guelph, Guelph, Ontario N1G 2W1, Canada}

\author{V.~Bildstein}
\affiliation{Department of Physics, University of Guelph, Guelph, Ontario N1G 2W1, Canada}

\author{I.~Dillmann}
\email[]{dillmann@triumf.ca}
\affiliation{TRIUMF, 4004 Wesbrook Mall, Vancouver, British Columbia V6T 2A3, Canada}
\affiliation{Department of Physics and Astronomy, University of Victoria, Victoria, British Columbia V8P 5C2, Canada}

\author{A.~Jungclaus}
\affiliation{Instituto de Estructura de la Materia, CSIC, E-28006 Madrid, Spain}

\author{C.~E.~Svensson}
\affiliation{Department of Physics, University of Guelph, Guelph, Ontario N1G 2W1, Canada}

\author{C.~Andreoiu}
\affiliation{Department of Chemistry, Simon Fraser University, Burnaby, British Columbia V5A 1S6, Canada}

\author{G.~C.~Ball}
\affiliation{TRIUMF, 4004 Wesbrook Mall, Vancouver, British Columbia V6T 2A3, Canada}

\author{N.~Bernier}
\affiliation{TRIUMF, 4004 Wesbrook Mall, Vancouver, British Columbia V6T 2A3, Canada}
\affiliation{Department of Physics and Astronomy, University of British Columbia, Vancouver, British Columbia V6T 1Z4, Canada}

\author{H.~Bidaman}
\affiliation{Department of Physics, University of Guelph, Guelph, Ontario N1G 2W1, Canada}

\author{P.~Boubel}
\affiliation{Department of Physics, University of Guelph, Guelph, Ontario N1G 2W1, Canada}

\author{C.~Burbadge}
\affiliation{Department of Physics, University of Guelph, Guelph, Ontario N1G 2W1, Canada}

\author{R.~Caballero-Folch}
\affiliation{TRIUMF, 4004 Wesbrook Mall, Vancouver, British Columbia V6T 2A3, Canada}

\author{M.~R.~Dunlop}
\affiliation{Department of Physics, University of Guelph, Guelph, Ontario N1G 2W1, Canada}

\author{L.~J.~Evitts}
\affiliation{TRIUMF, 4004 Wesbrook Mall, Vancouver, British Columbia V6T 2A3, Canada}
\affiliation{Department of Physics, University of Surrey, Guildford GU2 7XH, United
Kingdom}

\author{F.~Garcia}
\affiliation{Department of Chemistry, Simon Fraser University, Burnaby, British Columbia V5A 1S6, Canada}

\author{A.~B.~Garnsworthy}
\affiliation{TRIUMF, 4004 Wesbrook Mall, Vancouver, British Columbia V6T 2A3, Canada}

\author{P.~E.~Garrett}
\affiliation{Department of Physics, University of Guelph, Guelph, Ontario N1G 2W1, Canada}

\author{G.~Hackman}
\affiliation{TRIUMF, 4004 Wesbrook Mall, Vancouver, British Columbia V6T 2A3, Canada}

\author{S.~Hallam}
\affiliation{TRIUMF, 4004 Wesbrook Mall, Vancouver, British Columbia V6T 2A3, Canada}
\affiliation{Department of Physics, University of Surrey, Guildford GU2 7XH, United
Kingdom}

\author{J.~Henderson}
\affiliation{TRIUMF, 4004 Wesbrook Mall, Vancouver, British Columbia V6T 2A3, Canada}

\author{S.~Ilyushkin}
\affiliation{Department of Physics, Colorado School of Mines, Golden, Colorado 80401, USA}

\author{D.~Kisliuk}
\affiliation{Department of Physics, University of Guelph, Guelph, Ontario N1G 2W1, Canada}

\author{R.~Kr\"ucken}
\affiliation{TRIUMF, 4004 Wesbrook Mall, Vancouver, British Columbia V6T 2A3, Canada}
\affiliation{Department of Physics and Astronomy, University of British Columbia, Vancouver, British Columbia V6T 1Z4, Canada}

\author{J.~Lassen}
\affiliation{TRIUMF, 4004 Wesbrook Mall, Vancouver, British Columbia V6T 2A3, Canada}
\affiliation{Department of Physics and Astronomy, University of Manitoba, Winnipeg, Manitoba R3T 2N2, Canada}

\author{R.~Li}
\affiliation{TRIUMF, 4004 Wesbrook Mall, Vancouver, British Columbia V6T 2A3, Canada}

\author{E.~MacConnachie}
\affiliation{TRIUMF, 4004 Wesbrook Mall, Vancouver, British Columbia V6T 2A3, Canada}

\author{A.~D.~MacLean}
\affiliation{Department of Physics, University of Guelph, Guelph, Ontario N1G 2W1, Canada}

\author{E.~McGee}
\affiliation{Department of Physics, University of Guelph, Guelph, Ontario N1G 2W1, Canada}

\author{M.~Moukaddam}
\affiliation{TRIUMF, 4004 Wesbrook Mall, Vancouver, British Columbia V6T 2A3, Canada}

\author{B.~Olaizola}
\affiliation{Department of Physics, University of Guelph, Guelph, Ontario N1G 2W1, Canada}

\author{E.~Padilla-Rodal}
\affiliation{Universidad Nacional Aut\'onoma de M\'exico, Instituto de Ciencias Nucleares, AP 70-543, M\'exico City
04510, DF, M\'exico}

\author{J.~Park}
\affiliation{TRIUMF, 4004 Wesbrook Mall, Vancouver, British Columbia V6T 2A3, Canada}
\affiliation{Department of Physics and Astronomy, University of British Columbia, Vancouver, British Columbia V6T 1Z4, Canada}

\author{O.~Paetkau}
\affiliation{TRIUMF, 4004 Wesbrook Mall, Vancouver, British Columbia V6T 2A3, Canada}

\author{C.~M.~Petrache}
\affiliation{Centre de Sciences Nucl\'eaires et Sciences de la Mati\`ere, CNRS/IN2P3, Universit\'e Paris-Saclay, 91405 Orsay, France}

\author{J.~L.~Pore}
\affiliation{Department of Chemistry, Simon Fraser University, Burnaby, British Columbia V5A 1S6, Canada}

\author{A.~J.~Radich}
\affiliation{Department of Physics, University of Guelph, Guelph, Ontario N1G 2W1, Canada}

\author{P.~Ruotsalainen}
\affiliation{TRIUMF, 4004 Wesbrook Mall, Vancouver, British Columbia V6T 2A3, Canada}

\author{J.~Smallcombe}
\affiliation{TRIUMF, 4004 Wesbrook Mall, Vancouver, British Columbia V6T 2A3, Canada}

\author{J.~K.~Smith}
\affiliation{TRIUMF, 4004 Wesbrook Mall, Vancouver, British Columbia V6T 2A3, Canada}

\author{S.~L.~Tabor}
\affiliation{Department of Physics, Florida State University, Tallahassee, Florida 32306, USA}

\author{A.~Teigelh{\"o}fer}
\affiliation{TRIUMF, 4004 Wesbrook Mall, Vancouver, British Columbia V6T 2A3, Canada}
\affiliation{Department of Physics and Astronomy, University of Manitoba, Winnipeg, Manitoba R3T 2N2, Canada}

\author{J.~Turko}
\affiliation{Department of Physics, University of Guelph, Guelph, Ontario N1G 2W1, Canada}

\author{T.~Zidar}
\affiliation{Department of Physics, University of Guelph, Guelph, Ontario N1G 2W1, Canada}


\date{\today}

\begin{abstract}

The $\beta$-decay half-lives of $^{128\text{--}130}$Cd have been measured with the newly commissioned GRIFFIN $\gamma$-ray spectrometer at the TRIUMF-ISAC facility.
The time structures of the most intense $\gamma$-rays emitted following the $\beta$-decay were used to determine the half-lives of $^{128}$Cd and $^{130}$Cd to be $T_{1/2}= 246.2(21)$~ms and $T_{1/2}= 126(4)$~ms, respectively. The half-lives of the 3/2$^+$ and 11/2$^-$ states of $^{129}$Cd were measured to be $T_{1/2}(3/2^+)= 157(8)$~ms and $T_{1/2}(11/2^-)= 147(3)$~ms. The half-lives of the Cd isotopes around the $N=82$ shell closure are an important ingredient in astrophysical simulations to derive the magnitude of the second $r$-process abundance peak in the $A\sim130$ region. Our new results are compared with recent literature values and theoretical calculations.



\end{abstract}


\maketitle
%
%

\emph{Introduction} --- The $\beta$-decay properties (half-lives and $\beta$-delayed neutron-branching ratios) of nuclei below doubly-magic $^{132}_{\phantom{1}50}\text{Sn}_{82}^{}$ (i.e. $N\approx82$, $Z<50$) are key input parameters for any astrophysical $r$-process scenario as they play an important role in the formation and shape of the second abundance peak at $A\sim 130$~\cite{mumpower16}. This is despite the fact that the astrophysical site, or sites, where rapid neutron capture nucleosynthesis~\cite{bbfh57, cam57,CTT91,AGT07,TDK10} takes place remain(s) elusive.

In both the high- and low-entropy hot neutrino-driven wind scenarios,
the most important nuclei in this mass region are the $N=82$ isotones with $Z=40-50$ because the enhanced neutron binding energy compared to their isotopic neighbors leads to a barrier for the $r$-process reaction flow towards heavier masses. After the break-out of the $N=82$ shell, isotopes with $N=84$, 86 and 88 also become important, such as $^{134,136,138}$Sn, $^{133,135}$Ag, $^{132,134,136}$Cd, $^{131,133}$Rh, and $^{130}$Pd~\cite{mumpower16}. 

At the so-called ``waiting-point nuclei'', an accumulation of $r$-process material occurs (under given astrophysical conditions) and material can be transferred to the next elemental chain via $\beta$-decay. The half-lives of these waiting points thus determine how much material is accumulated, and hence, the amplitude and shape of the resulting $r$-process abundance peaks after decay back to stability. The prominent $r$-process abundance peaks at $A\sim 80$, 130, and 195 correspond to waiting-point isotopes at the closed neutron shells $N=50$, 82, and 126 where due to nuclear shell structure effects, the reaction flow is hindered. 




More neutron-rich ``cold'' $r$-process scenarios like neutron star mergers
\cite{freiburghaus99,korobkin12}, 
drive the reaction path towards the neutron dripline, into regions that will only be partially accessible to experiments at the new generation of radioactive beam facilities. Since most of the nuclei involved in $r$-process calculations are currently experimentally inaccessible, one has to rely heavily on the predictive power of theoretical models for the $\beta$-decay of these nuclei. The relative $r$-process abundances of nuclei around neutron shell-closures are particularly sensitive to the half-lives, and it is thus critical to have models that can accurately reproduce these decay properties. 

In particular, shell-model calculations for the waiting-point nuclei near the $N=82$ neutron shell closure~\cite{cuenca07,zhi13} have been performed by adjusting the quenching of the Gamow-Teller (GT) operator to reproduce the $^{130}$Cd half-life reported in Ref.~\cite{hannawald00}, and are known to yield systematically large values for the half-lives of other nuclei in the region~\cite{zhi13}. A new, shorter, half-life for $^{130}$Cd as measured by the EURICA collaboration~\cite{lorusso15} would resolve this discrepancy by scaling the GT quenching by a constant factor for all of the nuclei in this region.

Distinguishing between these discrepant half-life measurements for $^{130}$Cd \cite{hannawald00,lorusso15} is thus of critical importance since the as yet unknown half-lives of other $N=82$ waiting-point nuclei with $40 \le Z \le 44$ play a key role for the reproduction of the second abundance peak in $r$-process calculations. 



A recent experimental campaign with the EURICA detector~\cite{soderstrom13} at RIKEN measured the $\beta$-decay half-lives of 110 neutron-rich isotopes between Rb and Sn, among them, $^{128-130}$Cd~\cite{lorusso15}. While the previously reported half-life value for $^{128}$Cd of $T_{1/2}= 280(40)$~ms \cite{fogelberg88} was in agreement with the much more precise value of $T_{1/2}= 245 (5)$~ms reported in Ref.~\cite{lorusso15}, large discrepancies were found for $^{129,130}$Cd.

$^{129}$Cd $\beta$-decays from both a 3/2$^+$ and an 11/2$^-$ state but it is presently unknown which of the two states is the ground state and which is the isomeric state. In an experiment at ISOLDE, Arndt \emph{et al.}~\cite{arndt03,arndt09} measured the half-lives of both states using $\beta$-delayed neutrons. They reported $T_{1/2}= 104(6)$~ms for the ``11/2$^-$ ground-state'' and $T_{1/2}= 242(8)$~ms for the ``3/2$^+$ isomer'' but gave no explanation of how the ground state was assigned. 

The ground-state spin assignments for many neutron-rich odd-$A$ Cd isotopes have recently been confirmed via laser spectroscopy~\cite{yordanov13}. The odd-$A$ Cd isotopes $^{121,123,125,127}$Cd have a well-established ground-state spin of 3/2$^+$ but the exact position of the 11/2$^-$ isomer is not known for $^{125,127}$Cd. Shell-model calculations~\cite{taprogge15} suggest that this order is inverted at $^{129}$Cd compared to the lighter odd-$A$ Cd isotopes, however, there is no direct experimental evidence for this inversion. We thus label the two states only according to their spin and parity.

The recent measurements of the EURICA collaboration \cite{taprogge15} did not resolve the issue of the ground-state spin-parity assignment in $^{129}$Cd. However, the half-lives for the 3/2$^+$ and 11/2$^-$ states were determined separately via the $\gamma$-transitions at 1423 and 1586~keV in the daughter nucleus to be $T_{1/2}(3/2^+)= 146(8)$~ms, and via the $\gamma$-ray transitions at 359, 995, 1354, 1796, and 2156~keV to be $T_{1/2}(11/2^-)= 155(3)$~ms. These results are in clear contradiction with the previous measurements~\cite{arndt03,arndt09}. 

In the case of the $^{130}$Cd half-life, the value of 127(2)~ms reported in Ref.~\cite{lorusso15} also differed from the previously accepted value of 162(7)~ms~\cite{hannawald00} by more than 5$\sigma$. The measurement of Ref.~\cite{hannawald00} was performed with the same technique using $\beta$-delayed neutrons as the $^{129}$Cd measurements of Ref.~\cite{arndt03,arndt09}. In an earlier paper, the $^{130}$Cd half-life was reported as 195(35)~ms~\cite{kratz86}. 

In this paper we report an independent determination of the half-lives of $^{128-130}$Cd which, in general, confirm the recent EURICA results~\cite{lorusso15,taprogge15,tapPhD15}, but disagree with the previous measurements~\cite{arndt03,arndt09,hannawald00}. We report an improved precision for the $^{128}$Cd half-life, and revised half-lives for the two $\beta$-decaying states of $^{129}$Cd based on more detailed $\gamma$-ray spectroscopy.

\emph{Experiment} --- The half-lives of $^{128\text{-}130}$Cd were measured with the newly commissioned GRIFFIN $\gamma$-ray spectrometer~\cite{svensson14,rizwan16} at the TRIUMF-ISAC facility~\cite{dilling13}. Many of the nuclei in this neutron-rich region below doubly-magic $^{132}$Sn have complicated decay chains, including significant $\beta$-delayed neutron-emission branches, as well as the presence of $\beta$-decaying isomeric states. A measurement of the temporal distribution of characteristic $\gamma$-rays emitted from the excited states of the daughter nucleus following $\beta$-decay of the parent isotope is a powerful method to reduce the complex background contributions to the measurement. This method requires the use of a high-efficiency $\gamma$-ray spectrometer because of the low production rates and short half-lives.

The isotopes of interest were produced using a 500-MeV proton beam with 9.8-$\mu$A intensity from the TRIUMF main cyclotron incident on a UC$_x$ target. The ion-guide laser ion source (IG-LIS)~\cite{raeder14} was used to suppress surface-ionized isobars such as In and Cs, while the neutral Cd atoms of interest were extracted and selectively laser-ionized in a three-step excitation scheme. The Cd isotopes of interest were then accelerated to 28~keV, selected by a high-resolution mass separator and delivered to the GRIFFIN spectrometer. GRIFFIN is comprised of 16 high-purity germanium (HPGe) clover detectors~\cite{svensson14,rizwan16}. The radioactive ion beam (RIB) was implanted into an aluminized mylar tape of the moving tape collector at the mutual centers of SCEPTAR, an array of 20 thin plastic scintillators for tagging $\beta$ particles~\cite{ball05}, and GRIFFIN. The longer-lived background activity, either from isobaric contaminants in the beam or from daughters following the decay of the Cd isotopes, could be removed by moving the tape following a measurement. A typical cycle for the $^{128-130}$Cd runs consisted of a background measurement for 0.5~s, followed by a collection period (beam-on) with the beam being implanted into the tape for 10~s, followed by a collection period (beam-off) with the beam blocked by the ISAC electrostatic beam kicker downstream of the mass separator. The beam-off time consisted of a decay time of typically 2-3 half-lives, the movement of the tape for 1~s to a shielded position outside of the array, and the start of the new cycle with the background measurement. 





The high efficiency of the GRIFFIN array coupled with the SCEPTAR $\beta$ detector allowed for the sensitive detection of the $\gamma$-rays following the $\beta$-decay of interest. All of the analyses reported here were performed using add-back algorithms in which all of the detected energy in a clover within a 400-ns coincidence timing window was summed in order to increase the photopeak efficiency of GRIFFIN, as well as reduce the contribution of Compton background to the $\gamma$-ray spectrum~\cite{svensson14,rizwan16}.

\emph{Data Analysis} --- The data were analyzed using $\beta$-$\gamma$ coincidences requiring a $\beta$ particle to be detected in SCEPTAR within a coincidence window of 400~ns of a $\gamma$-ray detected in GRIFFIN, resulting in a strong suppression of room-background $\gamma$ rays. Cycles in which the RIB dropped out for a portion of the cycle were rejected in order to increase the signal to background ratio. In the case of $^{128}$Cd, this resulted in the removal of 30\% of the cycles, but only 1\% of the total data.




\emph{$^{128}$Cd Decay} --- Approximately 7~hrs of $^{128}$Cd data were collected with a beam intensity of $\sim$1000~pps. The 857-keV ($I_{\gamma,rel} = 95(10)\%$) and the 925-keV transition ($I_{\gamma,rel}= 12.4(12)\%$) in the daughter nucleus $^{128}$In~\cite{gokturk86} were used to determine the half-life. The strongest transition at 247~keV ($I_{\gamma,rel}= 100\%$) was not used because it is emitted from a 23(2)-$\mu$s isomeric state. The population of this isomer generally causes the emitted $\gamma$-ray to fall outside of the selected $\beta$-$\gamma$ time window of 400~ns. 

\begin{figure}[!t]
   \includegraphics[width=\linewidth]{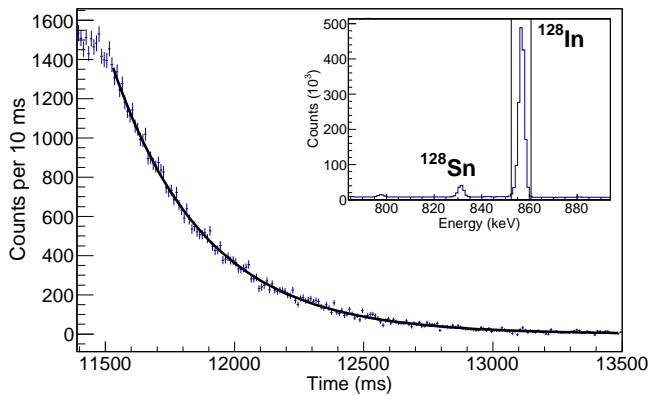}%
   \caption{Activity of the 857-keV $\gamma$-ray photopeak in $^{128}$In. The half-life of $^{128}$Cd from this transition is determined to be 245.8(21)~ms. The inset shows the gate on the 857-keV $\gamma$-ray as well as the strong 831.5-keV $\gamma$-ray following the $\beta$-decay of $^{128}$In. Note that the time represents the amount of time that has elapsed since the start of a cycle. Background data were collected for 1000~ms before the beam was turned on. The beam was turned off at 11.5~s.
   \label{fig:128hl}}
\end{figure}

The data were grouped into 10-ms bins and fitted with an exponential plus constant background, as shown for the 857-keV $\gamma$-ray in Fig.~\ref{fig:128hl}.
Sources of systematic uncertainties were investigated, including the re-binning of the data as well as a ``chop analysis''~\cite{grinyer05, laffoley13}. The chop analysis was performed by changing the fit region that was used in order to investigate rate-dependent effects. By starting the fit at a later time, rate-dependent effects such as pile-up~\cite{grinyer07} and dead-time are gradually reduced. If these effects are statistically significant, they can be seen as a correlation between the half-life and time of the first bin used.
The measured half-life did not change significantly as the first bin in the fit region was increased, nor did the measured half-life change as the last bin in the fit region was decreased. 

The data were also re-binned into 20~ms and 40~ms per bin with no statistically significant change in the fitted half-life. The half-lives deduced from the 857-keV and 925-keV $\gamma$-rays were 245.8(21)~ms and 257(11)~ms, respectively, while the fit to the sum of these two $\gamma$ rays resulted in a half-life of 246.2(21)~ms. This result is consistent with the previous measurements of 245(5)~ms~\cite{lorusso15} and 280(40)~ms~\cite{fogelberg88} and improves the precision of the $^{128}$Cd half-life by a factor of 2.4. 

\begin{figure}[!t]
   \includegraphics[width=\linewidth]{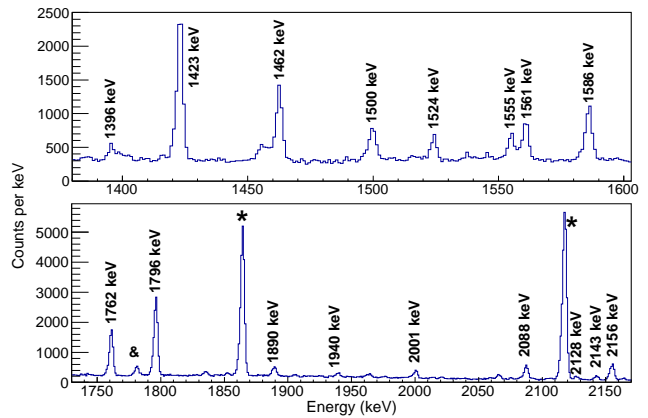}%
   \caption{A portion of the $\beta$-gated $\gamma$-ray energy spectrum for the $^{129}$Cd experiment. The strongest peaks in the spectrum are labeled. $\gamma$-rays emitted following the $\beta$-decay of $^{129}$Sn are labeled with a (*) and  $\gamma$-rays following the $\beta$-decay of $^{128}$Sn are labeled with a (\&). \label{fig:Cd129spectrum}}
\end{figure}
\begin{figure}[!t]
   \includegraphics[width=\linewidth,trim={0 6.5cm 0 0},clip]{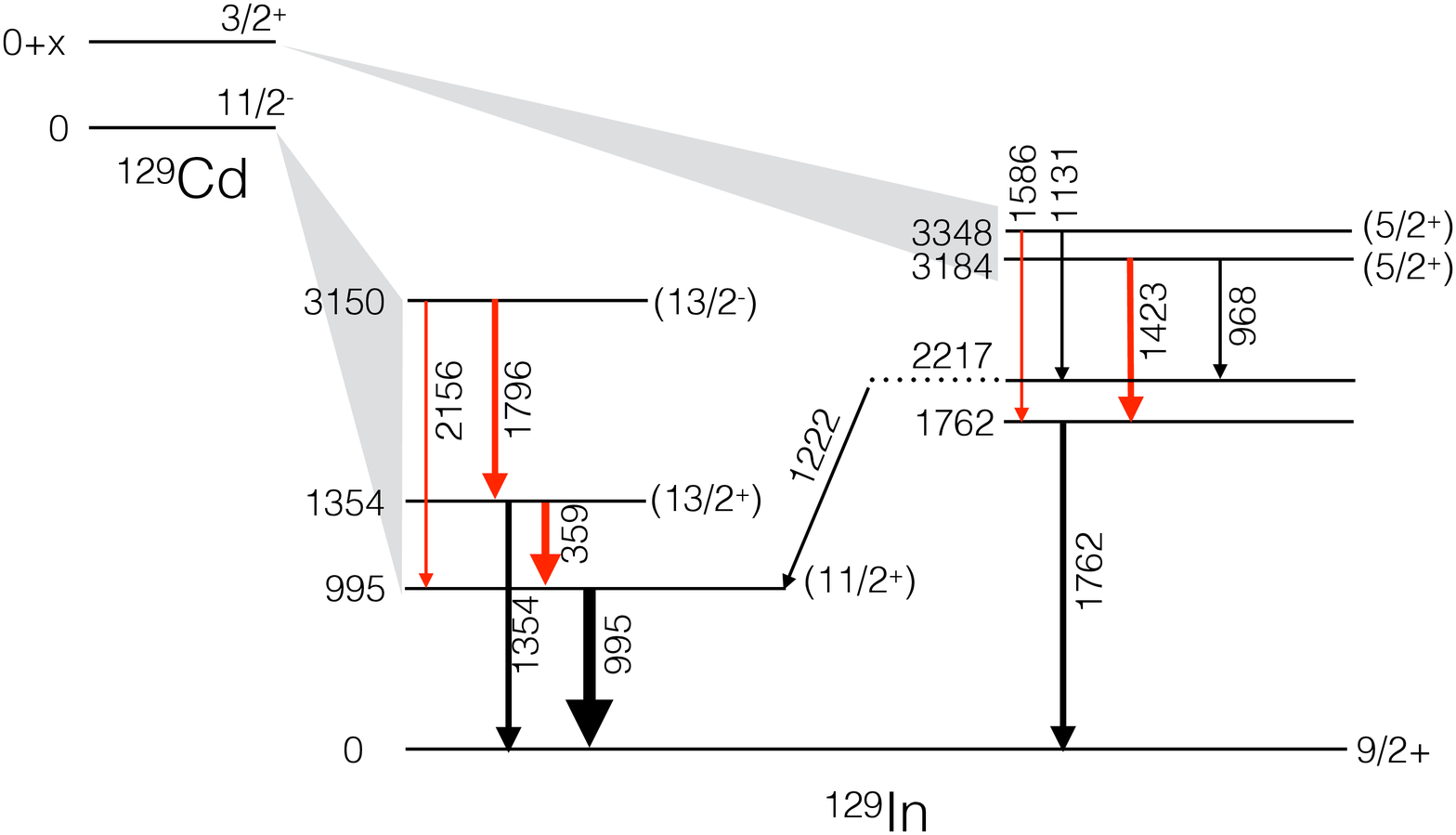}%
   \caption{(Color online) A portion of the $^{129}$Cd decay scheme (adapted from Ref.~\cite{taprogge15}) that shows the transitions relevant to the measurement of the half-life (see text for details). The $\gamma$-rays that were used in the half-life analysis are shown in red. Note that the ordering of the ground state and isomeric state in $^{129}$Cd is unknown; in this context ``$x$'' can be $<0$~keV.\label{fig:Cd129scheme}}
\end{figure}

\emph{$^{129}$Cd Decay} --- Approximately 13~hrs of $^{129}$Cd data were collected with a beam intensity of $\sim$250~pps.
The beam of $^{129}$Cd delivered to GRIFFIN consisted of both the ground state and the isomeric state. A portion of the $\gamma$-ray spectrum is shown in Fig.~\ref{fig:Cd129spectrum}, and the partial level scheme showing the important transitions for the measurement of the half-life is depicted in Fig.~\ref{fig:Cd129scheme}. 

The half-life of the $11/2^-$ state was deduced in Ref.~\cite{taprogge15} using the summed time distribution of the $\gamma$-transitions at 359, 995, 1354, 1796, and 2156~keV to be $T_{1/2}(11/2^-)= 155(3)$~ms. Similarly, the half-life of the $3/2^+$ state ($T_{1/2}(3/2^+)= 146(8)$~ms) was measured using the sum of the 1423~keV and 1586~keV $\gamma$ rays as they are known to be directly fed by the decay of the $3/2^+$ state in $^{129}$Cd. 

 
\begin{figure}[!t]
   \includegraphics[width=\linewidth]{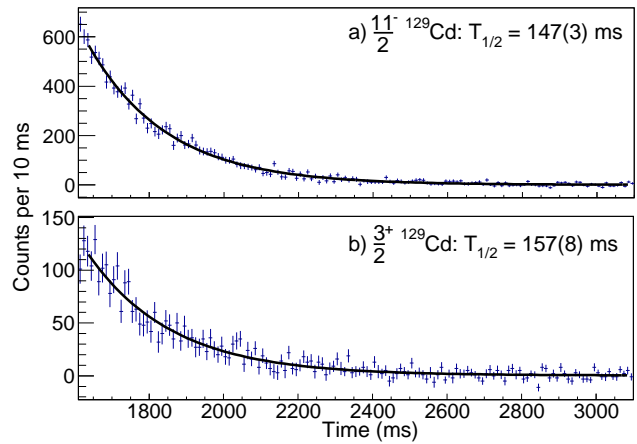}%
   \caption{a) Fitted activity of the $11/2^-$ state using the 358.9-, 1796.5- and 2155.6-keV $\gamma$-rays. b) Fitted activity of the $3/2^+$ state using the 1422.9- and 1586.2-keV $\gamma$-rays. Note that the time represents the amount of time that has elapsed since the start of a cycle.\label{fig:Cd129hl}}
\end{figure}

The high statistics of our measurements with GRIFFIN have made it possible to extract the half-life of the $11/2^-$ state from just the 359, 1796 and 2156~keV transitions, yielding $T_{1/2}(11/2^-)=147(3)$~ms. For~the half-life of the $3/2^+$ state, the two transitions at 1423 and 1586~keV were used resulting in $T_{1/2}(3/2^+)=157(8)$~ms~(Fig.~\ref{fig:Cd129hl}). A summary of the half-lives measured from each of the individual $\gamma$ rays is given in Table~\ref{tab:cd129halflives}. The same sources of systematic uncertainties as in the $^{128}$Cd analysis were also studied for $^{129}$Cd and found to have negligible influence compared to the statistical uncertainties. 

Unlike Ref.~\cite{taprogge15}, in this work we do not use the strong 995-keV $\gamma$ ray in the analysis of the half-life of the 11/2$^-$ state of $^{129}$Cd due to the previously observed feeding from the $3/2^+$ state of $^{129}$Cd via the 1222~keV transition in $^{129}$In~\cite{taprogge15}.
We also do not include the 1354-keV $\gamma$-ray in this analysis as it is contaminated by a $\gamma$-ray of the same energy from the decay of the 611(5)~ms ground state of $^{129}$In~\cite{gausemel04}. Based on the relative intensities of the observed $\gamma$-rays from $^{129}$In and $^{129}$Sn, we estimate that approximately 20\% of the total 1354-keV photopeak intensity in our experiment was from $^{129}$In decay, which, if~included, would bias the measured half-life to longer~values.

The half-lives for the $11/2^-$ and $3/2^+$ $\beta$-decaying states of $^{129}$Cd measured in this work agree with the general conclusion of~Ref.~\cite{taprogge15} that the half-lives of the two states are very similar, and do not differ by a factor of $\approx2$, as reported in Ref.~\cite{arndt03,kratz05,arndt09}. A direct comparison of the results from the individual $\gamma$-ray transitions between this work and those of Ref.~\cite{tapPhD15} is given in Table~\ref{tab:cd129halflives}. For statistical reasons, in Ref.~\cite{tapPhD15} the counts in the 1423- and 1586-keV photopeaks were summed, and the fit of the summed decay curve resulted in the published value of 146(8)~ms \cite{taprogge15} for the 3/2$^+$ state which is consistent with the value of 157(8)~ms reported here. The weighted average of these two independent measurements is 151.5(57)~ms.
For the half-life of the 11/2$^-$ state we do not average with Ref.~\cite{taprogge15}, but recommend the value of 147(3)~ms reported here due to the exclusion of contaminant $\gamma$-ray photopeaks in the current work.


\begin{table}[!t]
 \renewcommand{\thefootnote}{\roman{footnote}}
\caption{Half-life of $^{129}$Cd deduced from individual $\gamma$-rays in this work and the corresponding results from the measurements of Ref.~\cite{tapPhD15}  (see text for details).\label{tab:cd129halflives}}
 \begin{ruledtabular}
 \begin{tabular}{cllc}
  & \multicolumn{2}{c}{Half-life (ms)} & Parent state \\
$E_\gamma$ (keV) &  {this work} & {Ref.~\cite{tapPhD15}} & $J^\pi$ \\
 \hline
   358.9   &   148(3)   & 155.8(42) & 11/2$^-$\\
   1796.5  &   143(6)   & 157.9(99) & 11/2$^-$\\      
   2155.6  &   136(12)  & 144(24) & 11/2$^-$\\
   Weighted average & 146.5(26) & 155.8(38) & \\
   Summed value & 147(3) & 155(3)\footnote{The value published in Ref.~\cite{tapPhD15} includes the transitions at 995 and 1354~keV.} & \\
\hline
   1422.9  &   158(8)   &   & 3/2$^+$\\            
   1586.2  &   157(19)  &  & 3/2$^+$\\  
   Weighted average & 157.9(75) &  & \\
   Summed value & 157(8) & 146(8) & \\
\hline
   995.1$\footnote{$\gamma$-ray was not used in the current analysis due to a potential contamination between the $^{129}$Cd and $^{129}$Cd$^{\text{m}}$ decays.}$   &   151.8(23)  & 152.0(27) & 11/2$^-$ + 3/2$^+$\\
   1354.2\footnote{Doublet with a $\gamma$-ray from the decay of $^{129}$In.}  &   192(12)  & 158.6(81) & Doublet\\
 \end{tabular}
 \end{ruledtabular}
 \end{table}

\begin{figure}[!t]
   \includegraphics[width=\linewidth]{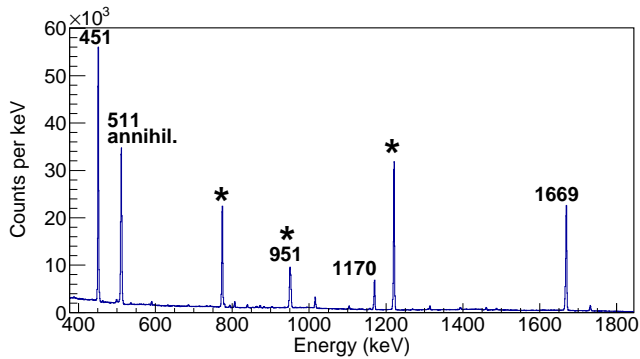}
   \caption{A portion of the $\beta$-gated $\gamma$-ray energy spectrum for the $^{130}$Cd experiment. The strongest peaks in the spectrum are labeled including the doublet at 951~keV. The three strong $\gamma$ rays at 451, 1170, and 1669~keV were used for the half-life analysis. $\gamma$ rays following the $\beta$ decay of $^{130}$In are labeled with a (*).\label{fig:130cdgamma}} 
\end{figure}

\emph{$^{130}$Cd Decay} --- For $^{130}$Cd approximately 38~hrs of data were collected with a beam intensity of 15-30~pps.
Figure~\ref{fig:130cdgamma} shows a portion of the $\beta$-coincident $\gamma$-ray spectrum obtained during the $^{130}$Cd experiment.

\begin{figure}[!t]
   \includegraphics[width=\linewidth]{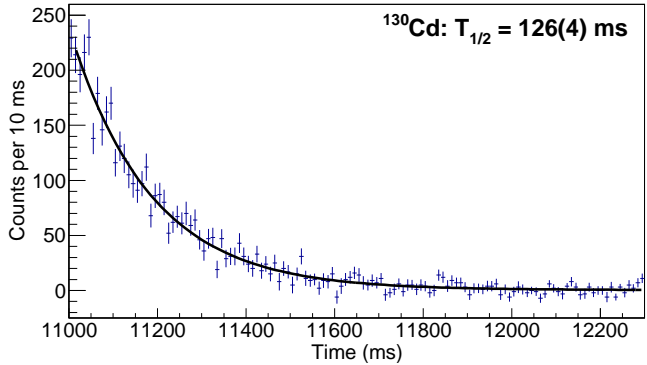}%
   \caption{Sum of the 451, 1170 and 1669~keV $\gamma$-ray time distributions. The half-life obtained from the fit is 126(4)~ms. Note that the time represents the amount of time that has elapsed since the start of a cycle.\label{fig:Cd130hl}}
\end{figure}

The 451.0- ($I_{\gamma,rel}= 88.6(36)\%$), 1170.3- ($I_{\gamma,rel}= 20.0(2)\%$), and 1669.2-keV ($I_{\gamma,rel}= 100\%$) $\gamma$-rays following the decay of $^{130}$Cd~\cite{dillmann03} were used to measure the half-life yielding 123(5)~ms, 138(20)~ms and 126(6)~ms, respectively. The transition at 951~keV ($I_{\gamma,rel}= 22.1(33)$\%)~\cite{dillmann03} was not used as it is a doublet with a $\gamma$ ray from the decay of $^{130}$In~\cite{fogelberg81}. Fitting the sum of the time distributions of these three $\gamma$ rays yields a half-life of 126(4)~ms for the decay of $^{130}$Cd (Fig.~\ref{fig:Cd130hl}), in excellent agreement with the value of 127(2)~ms recently reported in Ref.~\cite{lorusso15} and in strong disagreement with the previous half-life measurement of 162(7)~ms~\cite{hannawald00}. The study of systematic uncertainties was performed as discussed above, and did not reveal any statistically significant effects on the measured half-life.

\emph{Discussion and Conclusion} --- The half-lives of $^{128}$Cd, of the 11/2$^-$ and the 3/2$^+$ states of $^{129}$Cd, and of the $N=82$ isotope $^{130}$Cd were measured at TRIUMF-ISAC using the GRIFFIN $\gamma$-ray spectrometer. The $^{128}$Cd half-life measured in this work of 246.2(21)~ms is in excellent agreement with the previous measurement of Ref.~\cite{lorusso15}, but a factor of 2.4 more precise. The measured half-lives of the two known $\beta$-decaying states in $^{129}$Cd, 147(3)~ms for the $11/2^-$ state and 157(8)~ms for the $3/2^+$ state, are found to be similar, in agreement with the recent work of Ref.~\cite{taprogge15}, but in disagreement with the results of Refs.~\cite{arndt03,kratz05,arndt09}. We recommend the revised value for the $11/2^-$ state reported here rather than averaging with Ref.~\cite{taprogge15} due to the exclusion of potential contaminants in the current analysis.
Finally, the half-life of the $N=82$ waiting point nucleus $^{130}$Cd was measured to be 126(4)~ms, in excellent agreement with the value of 127(2)~ms reported in Ref.~\cite{lorusso15} but in strong disagreement with the measurements of 162(7)~ms and 195(35)~ms from Refs.~\cite{kratz86,hannawald00}. 

\begin{figure}[t!]
   \includegraphics[width=\linewidth]{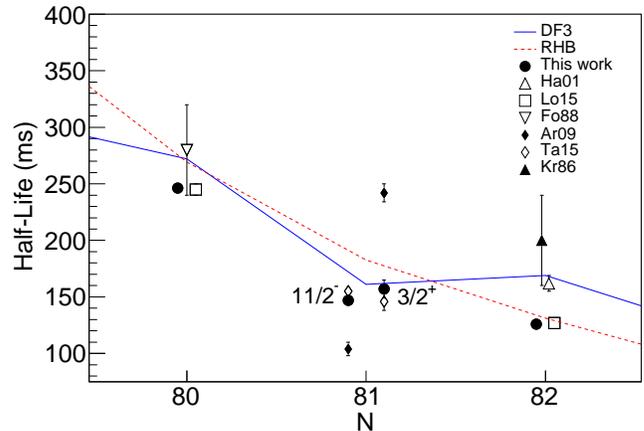}%
   \caption{(Color online) A comparison between the calculated half-lives using the DF3$+$CQRPA~\cite{borzov08} and RHB$+$RQRPA~\cite{marketin16} models to the measured half-lives of $^{128}$Cd, $^{129}$Cd and $^{130}$Cd. The measured half-lives in this work are in good agreement with Ref.~\cite{lorusso15,taprogge15} but do not agree with the previous measurements of Refs.~\cite{arndt03,arndt09,kratz05,hannawald00,fogelberg88}. Some error bars are not visible in the plot as they are smaller than the data points. Ha01:~\cite{hannawald00}, Lo15:~\cite{lorusso15}, Fo88:~\cite{fogelberg88}, Ar09:~\cite{arndt09}, Ta15~\cite{taprogge15}, Kr86~\cite{kratz86}.  \label{fig:df3compare}}
\end{figure}

The confirmation of the shorter half-life for the $N= 82$ isotope $^{130}$Cd has significant implications for nuclear structure calculations in this region, as well as for $r$-process nucleosynthesis simulations. As shown in Fig.~\ref{fig:df3compare}, the Fayans energy-density functional (DF3) $+$ continuum quasiparticle random phase approximation (CQRPA) model~\cite{borzov08} and the relativistic Hartee-Bogoliubov (RHB) $+$ relativistic quasiparticle random phase approximation (RQRPA)~\cite{marketin16} in general do a reasonable job of describing the systematic trend of the half-lives of neutron-rich Cd isotopes but slightly overestimate the absolute values. 

As discussed in Ref.~\cite{lorusso15}, the systematic overestimate of the half-lives for the $N=82$ isotones can be traced to the scaling of the Gamow-Teller quenching to the previously reported longer half-life for $^{130}$Cd~\cite{hannawald00}. Increasing the GT quenching factor from $q=0.66$ to $q=0.75$ in order to reproduce the shorter half-life of $^{130}$Cd reported in Ref.~\cite{lorusso15} and confirmed in the current work resolves this discrepancy. This directly affects the predicted half-lives for the yet unmeasured {$N=82$} isotones $^{127}$Rh, $^{126}$Ru and $^{125}$Tc. As demonstrated in Refs.~\cite{lorusso15,mumpower16}, the decrease in the calculated half-lives for these nuclei has a major influence on the shape of the rising wing of the $r$-process abundance peak at $A\sim130$.


This work has been partially supported by the Natural Sciences and Engineering Research Council of Canada (NSERC) and the Canada Research Chairs Program. I.D. and R.C.-F. are supported by NSERC Discovery Grants SAPIN-2014-00028 and RGPAS 462257-2014. A.J. acknowledges financial support by the Spanish Ministerio de Ciencia e Innovaci\'on under contract FPA2011-29854-C04 and the Spanish Ministerio de Econom\'ia y Competitividad under contract FPA2014-57196-C5-4-P. S.L.T acknowledges financial support from the U.S. National Science Foundation under contract NSF-14-01574. E.P.-R. acknowledges financial support from the DGAPA-UNAM under PASPA program. The GRIFFIN spectrometer was funded by the Canada Foundation for Innovation, TRIUMF, and the University of Guelph. TRIUMF receives federal funding via a contribution agreement with the National Research Council of Canada.

\end{document}